\documentclass[twocolumn]{aastex631}
\usepackage{xcolor}

\received{XXX}
\revised{YYY}
\accepted{ZZZ}

\shorttitle{Variability of OJ 287 from 2015 - 2023}
\shortauthors{Gupta et al.}

\graphicspath{{./}{}}

\begin{document}

\title{Quasi-simultaneous Optical Flux and Polarization Variability of the Binary Super Massive Black Hole Blazar OJ 287 from 2015 -- 2023: Detection of an Anti-correlation in Flux and Polarization Variability}

\author[0000-0002-9331-4388]{Alok C. Gupta}
\affiliation{Key Laboratory for Research in Galaxies and Cosmology, Shanghai Astronomical Observatory, Chinese Academy of Sciences, 80 Nandan Road, Shanghai 200030, People's Republic of China}
\affiliation{Aryabhatta Research Institute of Observational Sciences (ARIES), Manora Peak, Nainital -- 263001, India}
\email{acgupta30@gmail.com (ACG), pankaj.kushwaha@iisermohali.ac.in (PK), mvaltonen2001@yahoo.com (MJV)}

\author[0000-0001-6890-2236]{Pankaj Kushwaha}
\affiliation{Indian Institute of Science Education and Research (IISER) Mohali, Knowledge city, Sector 81, SAS Nagar, Manauli 140306, India}

\author[0000-0001-8580-8874]{Mauri J. Valtonen}
\affiliation{FINCA, University of Turku, FI-20014 Turku, Finland}
\affiliation{Tuorla Observatory, Department of Physics and Astronomy, University of Turku, FI-20014 Turku, Finland}

\author[0000-0003-4147-3851]{Sergey S.\ Savchenko}
\affiliation{Saint Petersburg State University, Universitetskaya nab. 7-9, St.  Petersburg, 199034, Russia}
\affiliation{Special Astrophysical Observatory, Russian Academy of Sciences, 369167, Nizhnil Arkhyz, Russia}
\affiliation{Pulkovo Observatory, St. Peterburg 196140, Russia}

\author[0000-0001-6158-1708]{Svetlana G. Jorstad}
\affiliation{Institute for Astrophysical Research, Boston University, 725 Commonwealth Avenue, Boston, MA 02215, USA}
\affiliation{Saint Petersburg State University, Universitetskaya nab. 7-9, St.  Petersburg, 199034, Russia}

\author[0000-0002-0643-7946]{Ryo Imazawa}
\affiliation{Department of Physics, Graduate School of Advanced Science and Engineering, Hiroshima University, 1-3-1 Kagamiyama, Higashi-Hiroshima, Hiroshima 739-8526, Japan}

\author[0000-0002-1029-3746]{Paul J. Wiita}
\affiliation{Department of Physics, The College of New Jersey, 2000 Pennington Rd., Ewing, NJ 08628-0718, USA}

\author[0000-0002-4455-6946]{Minfeng Gu}
\affiliation{Key Laboratory for Research in Galaxies and Cosmology, Shanghai Astronomical Observatory, Chinese Academy of Sciences, 80 Nandan Road, Shanghai 200030, People's Republic of China}

\author[0000-0001-7396-3332]{Alan P. Marscher}
\affiliation{Institute for Astrophysical Research, Boston University, 725 Commonwealth Avenue, Boston, MA 02215, USA}

\author[0000-0002-8366-3373]{Zhongli Zhang}
\affiliation{Shanghai Astronomical Observatory, Chinese Academy of Sciences, Shanghai 200030, People's Republic of China}
\affiliation{Key Laboratory of Radio Astronomy and Technology, Chinese Academy of Sciences, A20 Datun Road, Chaoyang District, Beijing 100101, People's Republic of China}

\author[0000-0002-0766-864X]{Rumen Bachev}
\affiliation{Institute of Astronomy and National Astronomical Observatory, Bulgarian Academy of Sciences, 72 Tsarigradsko shosse Blvd., 1784 Sofia, Bulgaria}

\author[0000-0002-7262-6710]{G. A. Borman}
\affiliation{Crimean Astrophysical Observatory RAS, P/O Nauchny, 298409, Russia}

\author[0000-0002-6629-8490]{Haritma Gaur}
\affiliation{Aryabhatta Research Institute of Observational Sciences (ARIES), Manora Peak, Nainital -- 263001, India}

\author[0000-0002-3953-6676]{T. S.\ Grishina}
\affiliation{Saint Petersburg State University, Universitetskaya nab. 7-9, St.  Petersburg, 199034, Russia}

\author[0000-0002-6431-8590]{V. A. Hagen-Thorn}
\affiliation{Saint Petersburg State University, Universitetskaya nab. 7-9, St.  Petersburg, 199034, Russia}

\author[0000-0001-9518-337X]{E. N.\ Kopatskaya}
\affiliation{Saint Petersburg State University, Universitetskaya nab. 7-9, St.  Petersburg, 199034, Russia}

\author[0000-0002-4640-4356]{V. M.\ Larionov}\thanks{Deceased}
\affiliation{Saint Petersburg State University, Universitetskaya nab. 7-9, St.  Petersburg, 199034, Russia}
\affiliation{Pulkovo Observatory, St. Peterburg 196140, Russia}

\author[0000-0002-2471-6500]{E. G. Larionova}
\affiliation{Saint Petersburg State University, Universitetskaya nab. 7-9, St.  Petersburg, 199034, Russia}

\author[0000-0002-0274-1481]{L. V. Larionova}
\affiliation{Saint Petersburg State University, Universitetskaya nab. 7-9, St.  Petersburg, 199034, Russia}

\author[0000-0002-9407-7804]{D. A. Morozova}
\affiliation{Saint Petersburg State University, Universitetskaya nab. 7-9, St.  Petersburg, 199034, Russia}

\author{T. Nakaoka}
\affiliation{Hiroshima Astrophysical Science Center, Hiroshima University, 1-3-1 Kagamiyama, Higashi-Hiroshima, Hiroshima 739-8526, Japan}

\author{A. Strigachev}
\affiliation{Institute of Astronomy and National Astronomical Observatory, Bulgarian Academy of Sciences, 72 Tsarigradsko shosse Blvd., 1784 Sofia, Bulgaria}

\author[0000-0002-9907-9876]{Yulia V. Troitskaya}
\affiliation{Saint Petersburg State University, Universitetskaya nab. 7-9, St.  Petersburg, 199034, Russia}

\author[0000-0002-4218-0148]{I. S. Troitsky}
\affiliation{Saint Petersburg State University, Universitetskaya nab. 7-9, St.  Petersburg, 199034, Russia}

\author[0000-0002-7375-7405]{M. Uemura}
\affiliation{Hiroshima Astrophysical Science Center, Hiroshima University, 1-3-1 Kagamiyama, Higashi-Hiroshima, Hiroshima 739-8526, Japan}

\author[0000-0002-8293-0214]{A. A. Vasilyev}
\affiliation{Saint Petersburg State University, Universitetskaya nab. 7-9, St.  Petersburg, 199034, Russia}

\author[0000-0001-6314-0690]{Z. R. Weaver}
\affiliation{Institute for Astrophysical Research, Boston University, 725 Commonwealth Avenue, Boston, MA 02215, USA}

\author{A. V. Zhovtan}
\affiliation{Crimean Astrophysical Observatory RAS, P/O Nauchny, 298409, Russia}

\begin{abstract}
\noindent
We study the optical flux and polarization variability of the binary black hole blazar OJ 287 using quasi-simultaneous observations from 2015 -- 2023 carried out using telescopes in the USA, Japan, Russia, Crimea and Bulgaria. This is one of the most extensive quasi-simultaneous optical flux and polarization variability studies of OJ 287. OJ 287 showed large amplitude, $\sim$ 3.0 magnitude flux variability, large changes of $\sim$ 37\%  in degree of polarization, and a large swing of $\sim$ 215$^{\circ}$ in the angle of the electric vector of polarization. During the period of observation, several flares in flux were detected. Those flares are correlated with rapid increase in degree of polarization and swings in electric vector of polarization angle. A peculiar behavior of anti-correlation between flux and polarization degree, accompanied by a nearly constant polarization angle, was  detected from JD 2458156 to JD 2458292.  We briefly discuss some  explanations for the flux and polarization variations observed in OJ 287.
\end{abstract}

\section{Introduction} \label{sec:introduction}
\noindent
OJ 287 is an archetypal blazar of the BL Lacertae class at a redshift of 0.306 \citep{1985PASP...97.1158S}. In its optical spectra, the lines appear only weakly above the non-thermal continuum, but are visible at most times \citep[for spectral surveys, see][]{2010A&A...516A..60N,2021Galax..10....1V}. Very high energy $\gamma-$ray emission $>$100 GeV was detected from OJ 287 with VERITAS ({\it Very Energetic Radiation Imaging Telescope Array System}) \citep{2017ATel10051....1M} and so it is listed as a TeV emitting blazar in the TeV catalogue\footnote{\url{http://tevcat.uchicago.edu/}}. Historically, it is one of the most variable extragalactic sources at optical and radio bands, exhibiting variations in both flux and polarization \citep{1987A&A...184...57V}. OJ 287 is among a few active galactic nuclei (AGN), which is understood to host an in-spiraling supermassive black hole (SMBH) binary system and hence is a special candidate to emit nano-hertz
gravitational waves
\citep[e.g.,][and references therein]{2021Galax..10....1V,2023MNRAS.521.6143V}. \\
\\
Being one of the most optically bright AGNs, its favorable location close to the ecliptic has led to optical data having been collected over a very extended period -- more than a century, since 1888. Using the available early subset of this data, \citet{1988ApJ...325..628S} discerned for the first time that the blazar shows a period of $\sim$ 12 yrs and using the similarity of flare profile with simulations, hypothesized it to be a binary SMBH. The model successfully predicted the next outburst as well as the expected tidally-induced, second peak, $\sim$ 1.2 yr later \citep{1996A&A...305L..17S,1996A&A...315L..13S}. The denser monitoring cadence during this period, however, revealed much sharper flare profiles, leading to a drastically refined model with flares arising out of the impact of secondary on the primary's accretion disk \citep{1996ApJ...460..207L}. Alternative hypotheses argue the recurrent flares arise from jet precession or a combination of jet and accretion disk emission \citep[e.g.][]{1997ApJ...478..527K,2023ApJ...951..106B}. Comparison of observational features explored to date (timing, spectral, and polarization) to those expected in these models, largely favor the disk impact scenario \citep{2020Galax...8...15K,2018MNRAS.473.1145K} yet the case is not ironclad. Given our incomplete understanding of accretion physics and compact sources, a persistent effort across the EM spectrum is crucial to eventual elimination of any lingering doubts \citep[][and references therein]{2023Galax..11...82V}. \\
\\
Along with BL Lacertae itself, OJ 287 was a prototype BL Lac object used for the observational characterization of BL Lac source properties. Thus its optical and near infra-red (NIR) flux and polarization properties have been extensively explored on diverse timescales \citep[e.g.,][and references therein]{1984MNRAS.211..497H,1988A&A...190L...8K,1991AJ....101.2017S,1996A&A...305L..17S,1996A&A...315L..13S,2017MNRAS.465.4423G,2019AJ....157...95G,2022ApJS..260...39G}. Optical and NIR polarization studies have reported frequency-dependent polarization \citep{1984MNRAS.211..497H,1988A&A...190L...8K} and led to the proposal that two-emission regions were important \citep[see also][]{2010MNRAS.402.2087V}. In optical bands and on intra-night timescales frequency-dependent polarization behavior has been reported with an anti-correlation between flux and polarization degree (PD) \citep{1991AJ....101.2017S}. \citet{2011ARep...55.1078B} analyzed four-color optical photometric data and R-band polarimetric data of the blazar OJ 287 over 5 yr (2005 -- 2009)
and employed a model with one constant, and several variable, sources of polarized radiation. They concluded that the observed variability of the polarization and colors  in the optical range could be explained with a model consisting of the constant source with a degree of polarization $\approx$10\% and position angle of polarization PA$\sim$162$^{\circ}$, and 4$\pm$2 variable sources with similar degrees of polarization and random PAs.\\
\\
The theory that a binary SMBH system is present in OJ 287 proposed that the $\sim12$-yr recurrent optical flares are of thermal bremsstrahlung origin \citep{1996ApJ...460..207L}. Hence polarization is one of the key observables for  discerning the nature of the flare, as it is expected to decrease during the impact flares because thermal emission has no polarization. The first evidence for this already was available from 1983, where the expected trend was seen \citep{1985AJ.....90.1184S}. During the rising part of the flares in 1990s polarization observations were not available and this was again the case in 2005. In 2005 the first polarization measurement was carried out after the peak flux \citep{2010MNRAS.402.2087V}. Therefore the next real test, since 1983, of the unpolarized nature of the impact flares came in 2007 when it was shown that the flare arose only from  the unpolarized component \citep{2008Natur.452..851V,2011AcPol..51f..76V}. The next opportunity to study the polarization of an impact flare came in 2015, when, in addition to demonstrating the unpolarized nature of the optical emission, it was shown that there was no X-ray counterpart to the flare \citep{2016ApJ...819L..37V} further supporting its thermal nature. \\
\\
The spectral slope over a wide range of frequencies was measured during the 2005 flare and the expected thermal bremmstrahlung spectral index of $\alpha_\nu\sim-0.2$ was found \citep{2012MNRAS.427...77V}. It is quite different from the usual spectral index of $\alpha_\nu\sim-1.3$ over the same spectral range \citep{2018A&A...610A..74K} which is typical for non-thermal synchrotron emission. The spectral index $\alpha_\nu\sim-0.2$ was confirmed during the 2015 and 2019 flares \citep{2020ApJ...894L...1L,2021Galax..10....1V}. The complete change in the nature of emission in the impact flare as compared with the usual out-of-flare emission excludes the possibility that the flares are somehow related to Doppler boosting variations in a turning jet \citep{1998MNRAS.293L..13V,2004ApJ...615L...5R}. \\
\\
Even though the nature of radiation during impact flares has now been well studied, there still remain questions. The second component after the bremsstrahlung peak is highly polarized \citep{1985AJ.....90.1184S,2010MNRAS.402.2087V,2016ApJ...819L..37V} which indicates that here we receive non-thermal emission of the impact flare in OJ~287 \citep{2019ApJ...882...88V}. Further flares associated with tidal increase of accretion flow were expected in 2016/2017 and 2020 \citep{1997ApJ...484..180S,2013ApJ...764....5P,2021Galax..10....1V}. A so-called precursor flare also falls into the time period studied here \citep{2013ApJ...764....5P}. In addition, the long range evolution of the polarization behaviour has been described by models which associate the polarization variability to the wobble of the jet, caused by the companion, and the resulting changes in the jet viewing angle \citep{2012MNRAS.427...77V,2013A&A...557A..28V,2021MNRAS.503.4400D,2021Galax..10....1V}.\\
\\
Here we report our dense quasi-simultaneous monitoring of the optical flux and polarization behavior of the source between 2015 -- 2023. These observations are part of our ongoing optical and multi-wavelength project to densely monitor the behavior of OJ 287 \citep{2017MNRAS.465.4423G,2019AJ....157...95G,2018MNRAS.473.1145K,2021ApJ...921...18K}. This interval  corresponds to much of one cycle of  the observationally-favored binary SMBH model of \citet[see \citet{1996ApJ...460..207L}]{2018ApJ...866...11D}. The next section details the optical monitoring and reduction procedure, followed by analysis and results in section \S\ref{sec:results}. A discussion and a summary are provided in sections \S\ref{sec:discuss} and \S\ref{sec:sum}, respectively.

\begin{figure*}
    \centering
    \includegraphics[scale=0.9]{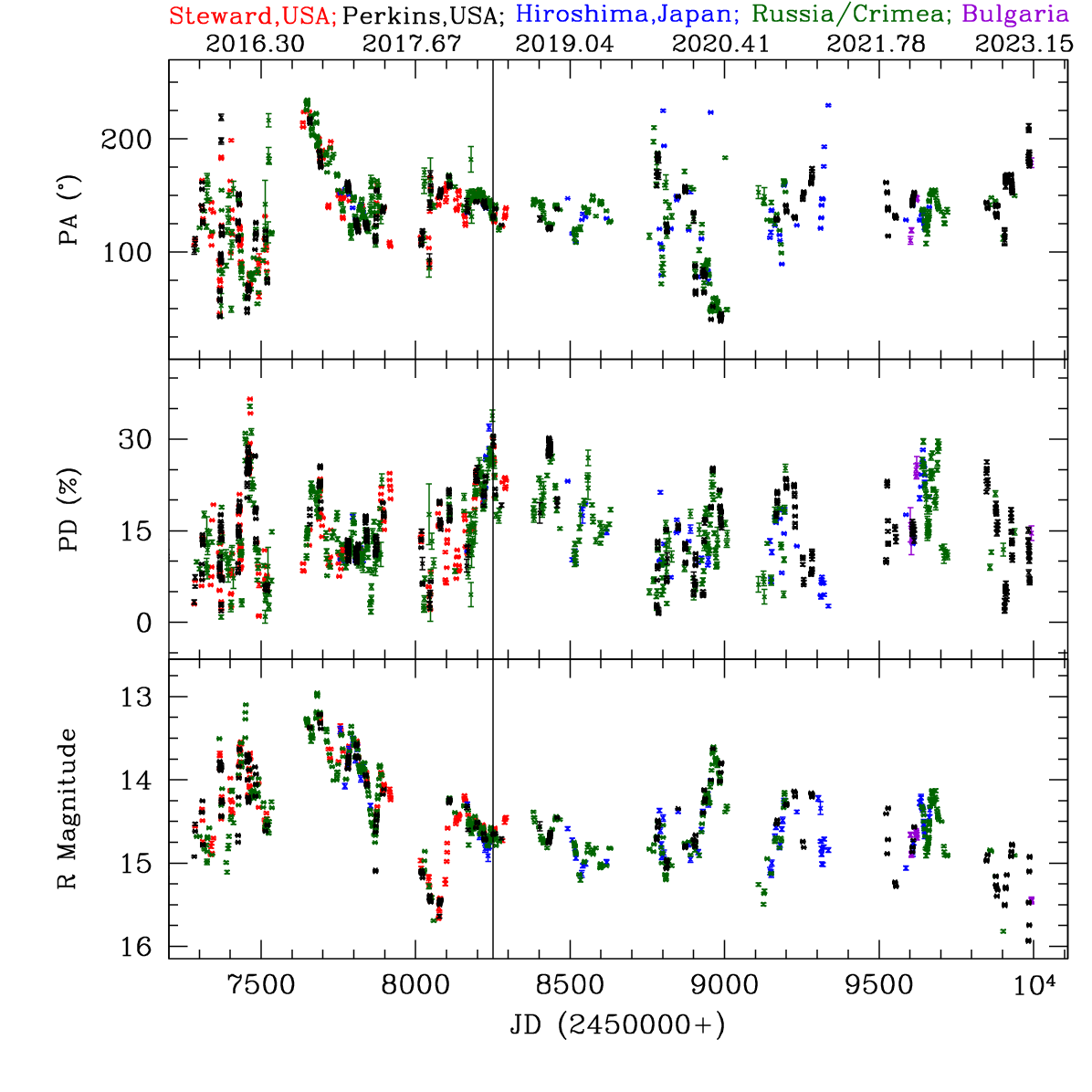}
    \caption{Simultaneous optical flux and polarization light curves of OJ 287 from 2015 to 2023, with the R magnitudes at the bottom and the PD and PA in the middle and upper panels, respectively.
}
    \label{fig:LC}
\end{figure*}

\section{Data Collection and Reduction}
\noindent
Optical R-band photopolarimetric observations of the blazar OJ 287 were performed during 2017 March 7 to 2022 March 26 with the Hiroshima Optical and Near-InfraRed camera (HONIR) \citep{2014SPIE.9147E..4OA} which is installed on the ``Kanata" 1.5m telescope, Hiroshima, Japan. Calculation of the PD,  PA, and their errors were done from Stokes parameters obtained from four exposures at 0.0$^{\circ}$, 45$^{\circ}$.0, 22$^{\circ}$.5, and 67$^{\circ}$.5 positions of the half-wave plate in each exposure \citep{1999PASP..111..898K}. We obtained the offset angle from observation of strongly polarized stars (BD+64d106, BD+59d389) and depolarization correction was performed from the wire grid star observation. We confirmed the instrumental polarization was negligible ($<$ 0.2\%) from observation of an unpolarized star (HD14069). For the photometric observations, standard reduction procedures were adopted and we calculated the magnitude based on the Pan-STARRS1 catalog after conversion of the filter system. The photometry of the blazar was done using standard CCD image reduction procedures. Detailed descriptions of observations, photometric and polarimetric data analyses made with this telescope and instrumentation are described in \citet{2011PASJ...63..639I}. The data from the Kanata telescope are plotted in blue symbols in the  different panels of Figure \ref{fig:LC}, which give the photometric magnitude, degree of polarization, and polarization angle from bottom to top.\\
\\
Optical photometric and polarimetric data of OJ 287 were obtained from 2015 September 25 to 2022 December 24 using the St.\ Petersburg University, Russia 40-cm LX-200 telescope and the Crimean observatory 70-cm AZT-8 telescope under regular monitoring of a sample of blazars within the framework of a GASP/WEBT (GLAST-AGILE Support Program/Whole Earth Blazar Telescope) project. We have used differential aperture photometry to obtain photometric measurements with standard stars 1, 2, and 3 in VRI from \citet{1996A&AS..116..403F} and in B from \citet{1985AJ.....90.1184S}. Detailed description of our data analyses are provided in \citet{2008A&A...492..389L}. The data from the St.\ Petersburg and Crimea telescopes are plotted in green in the different panels of Figure\ \ref{fig:LC}. \\
\\
Optical flux and polarimetric observations of OJ 287 were carried out from 2015 September 18 to 2023 February 12 at the Perkins telescope of the Perkins Telescope Observatory (Flagstaff, AZ, USA) \citep{2010ApJ...715..362J}. They were obtained with the PRISM camera\footnote{\url{https://www.bu.edu/prism/}} which possesses a polarimeter with a rotating half-waveplate. A polarization observation consists of a series of 3--5 Stokes Q and U measurements for a given object. Each series consists of four measurements at instrumental position angles 0$^{\circ}$, 45$^{\circ}$, 90$^{\circ}$, and 135$^{\circ}$ of the waveplate. We use field stars to perform both interstellar and instrumental polarization corrections. We use unpolarized calibration stars from \citet{1992AJ....104.1563S} to check the instrumental polarization, which is usually within 0.3\%--0.5\%, and polarized stars from the same paper to calibrate the polarization position angle. These data are plotted in black in Figure\ \ref{fig:LC}. \\
\\
Additional optical R band photometric and polarimetric data of OJ 287 from 2022 January 21 to 2023 February 17 were taken using the 0.6-m telescope of the Belogradchik Observatory, Bulgaria. A double-barrel filter set from FLI was installed in the Belogradchik 0.6-m telescope, enabling the combination of polarimetric and UBVRI filters to perform a linear polarimetry of the blazar in the chosen optical band. The PD and PA were obtained using three polarimetric filters that were orientated at 0-180, 60-240, and 120-300 degrees, respectively. Details about this data analysis are provided in \citet{2023MNRAS.522.3018B}. The data from Belogradchik Observatory telescope are plotted in violet in Figure\ \ref{fig:LC}. \\
\\
In addition to the above data that we collected, we also employed optical R band photometric and polarimetric data obtained from the spectro-polarimeter at Steward Observatory, University of Arizona, USA, on OJ 287 from 2015 September 18 to 2018 June 22  taken from the public archive\footnote{\url{http://james.as.arizona.edu/~psmith/Fermi/DATA/Objects/}}. The details of this observational program and analysis procedures are explained in \citet{2009arXiv0912.3621S}. The Steward Observatory photometric magnitude, degree of polarization, and polarization angle of OJ 287 are plotted in red in Figure \ref{fig:LC}. \\
\\
We note that during periods of overlap for measurements taken at different observatories the magnitude and fractional polarization measurements agree very well. We noticed  occasional ambiguity in the measurements of the polarization angles and corrected them by  $\pm \rm{n}\pi$, where n = 1, 2, .....

\begin{figure*}
\centering
\includegraphics[width=0.45\textwidth,height=0.45\textheight]{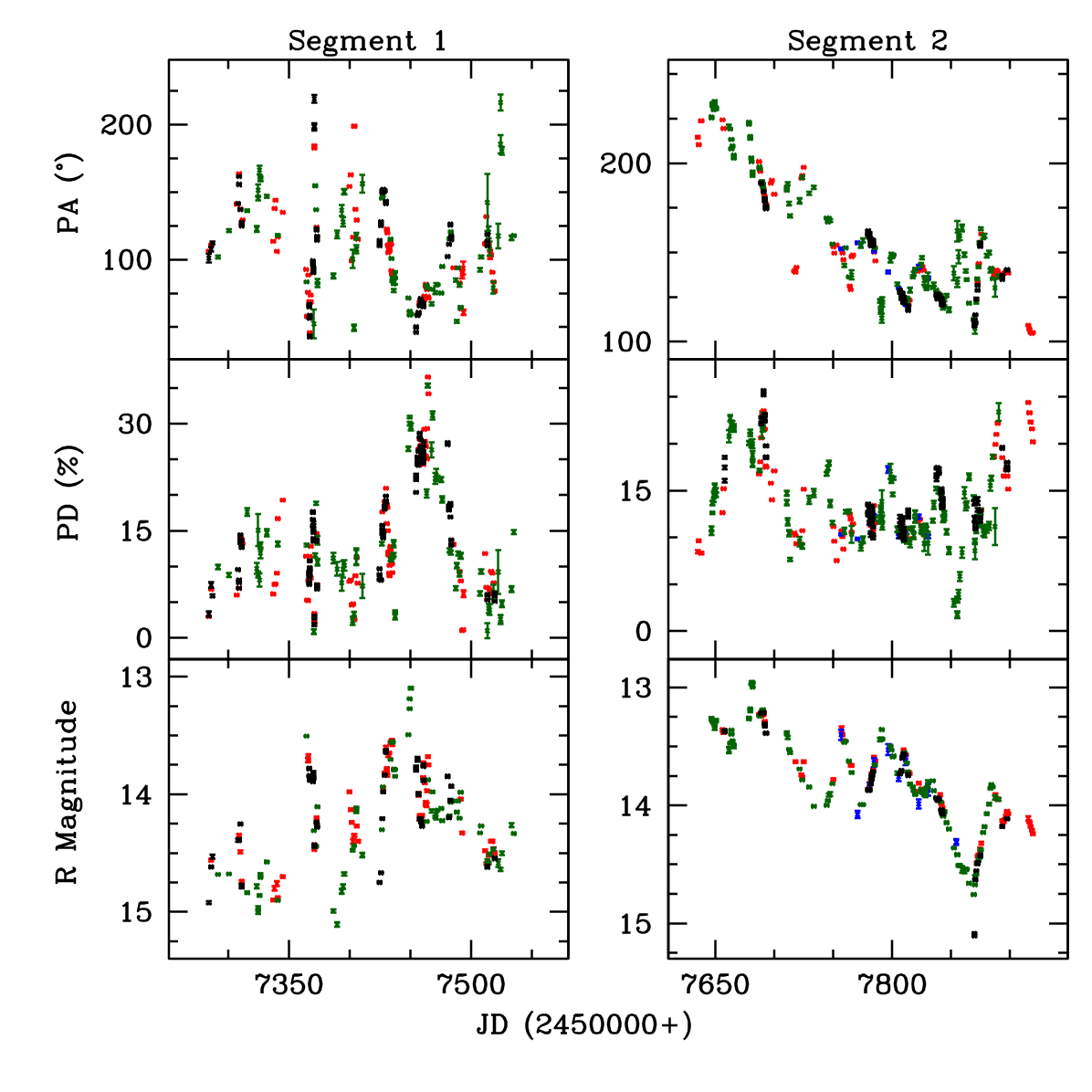}
\includegraphics[width=0.45\textwidth,height=0.45\textheight]{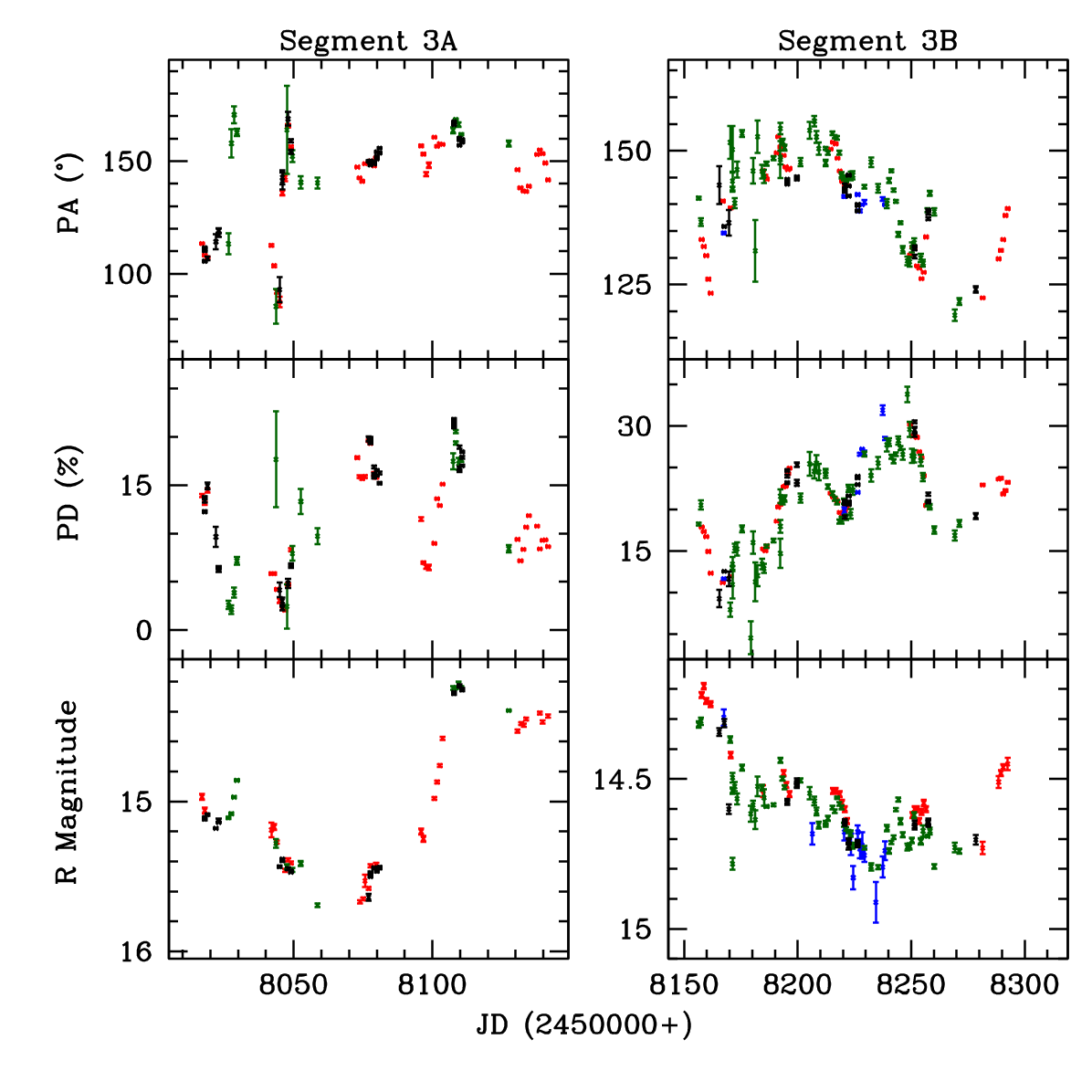}

\includegraphics[width=0.45\textwidth,height=0.45\textheight]{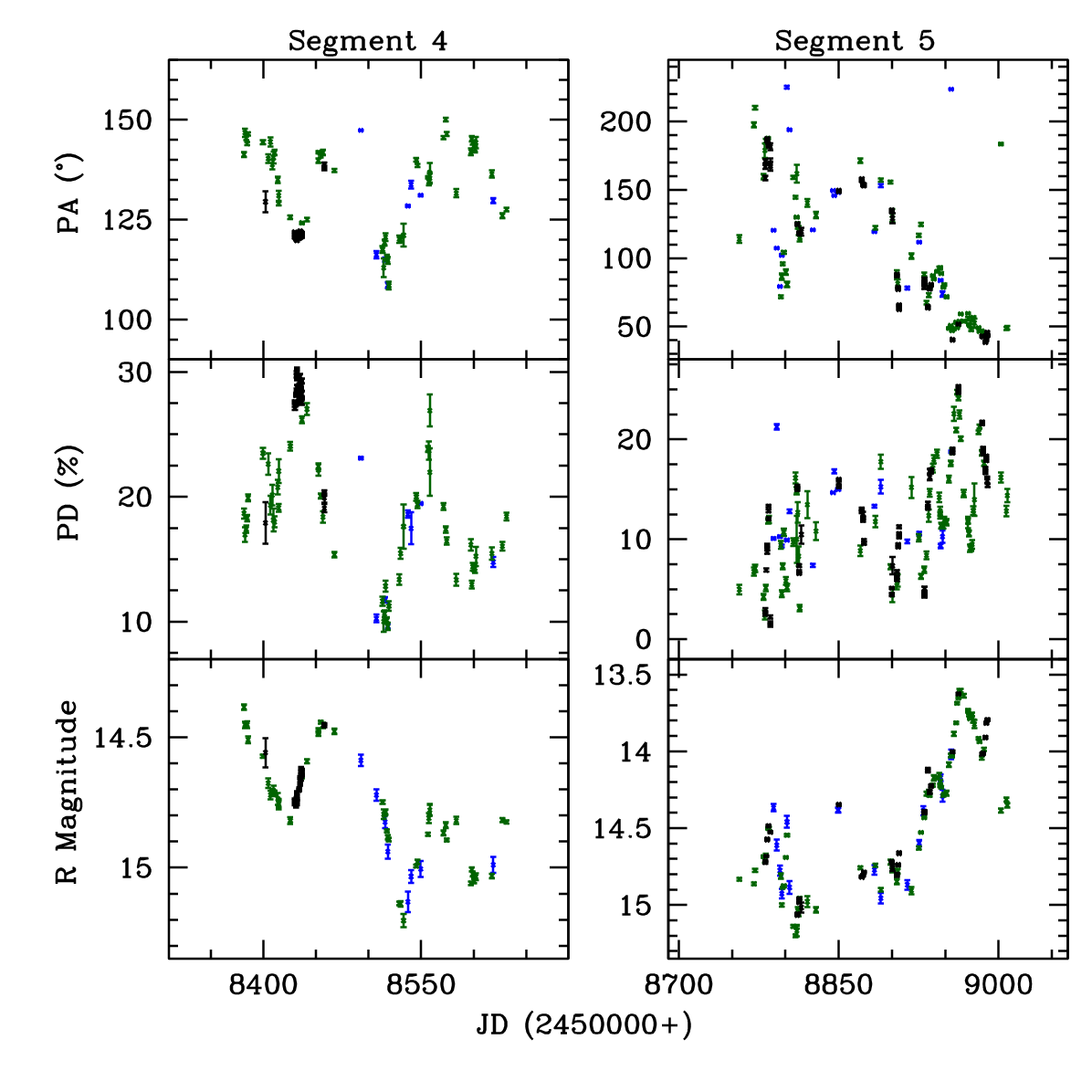}
\includegraphics[width=0.45\textwidth,height=0.45\textheight]{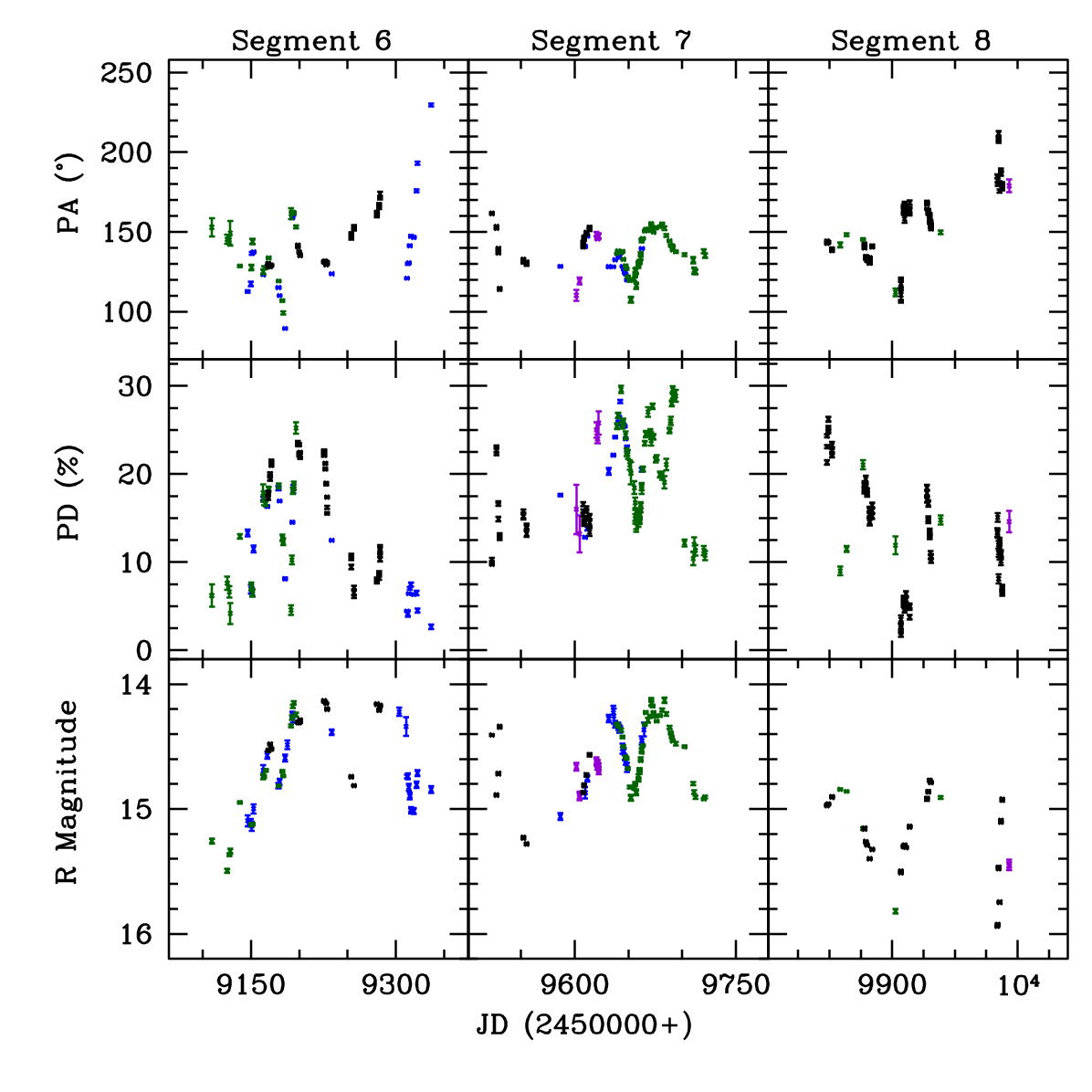}
\caption{Enlargements of portions of the observations shown in Figure \ref{fig:LC}, divided into eight segments in which segment 3 is divided into two sub-segments. Symbols of different colors are the same as in Figure \ref{fig:LC}.}
\label{fig:LC2}
\end{figure*}

\section{Results}\label{sec:results}
\noindent
The simultaneous optical photometric and polarimetric behavior of the blazar OJ 287 is displayed in Figure \ref{fig:LC}. On visual inspection, we noticed that the  flux and polarization degree are generally correlated. But on one occasion an anti-correlation between optical flux and PD is seen and is marked by a continuous line in Figure \ref{fig:LC}.
  
\begin{figure}
    \centering
    \includegraphics[scale=0.42]{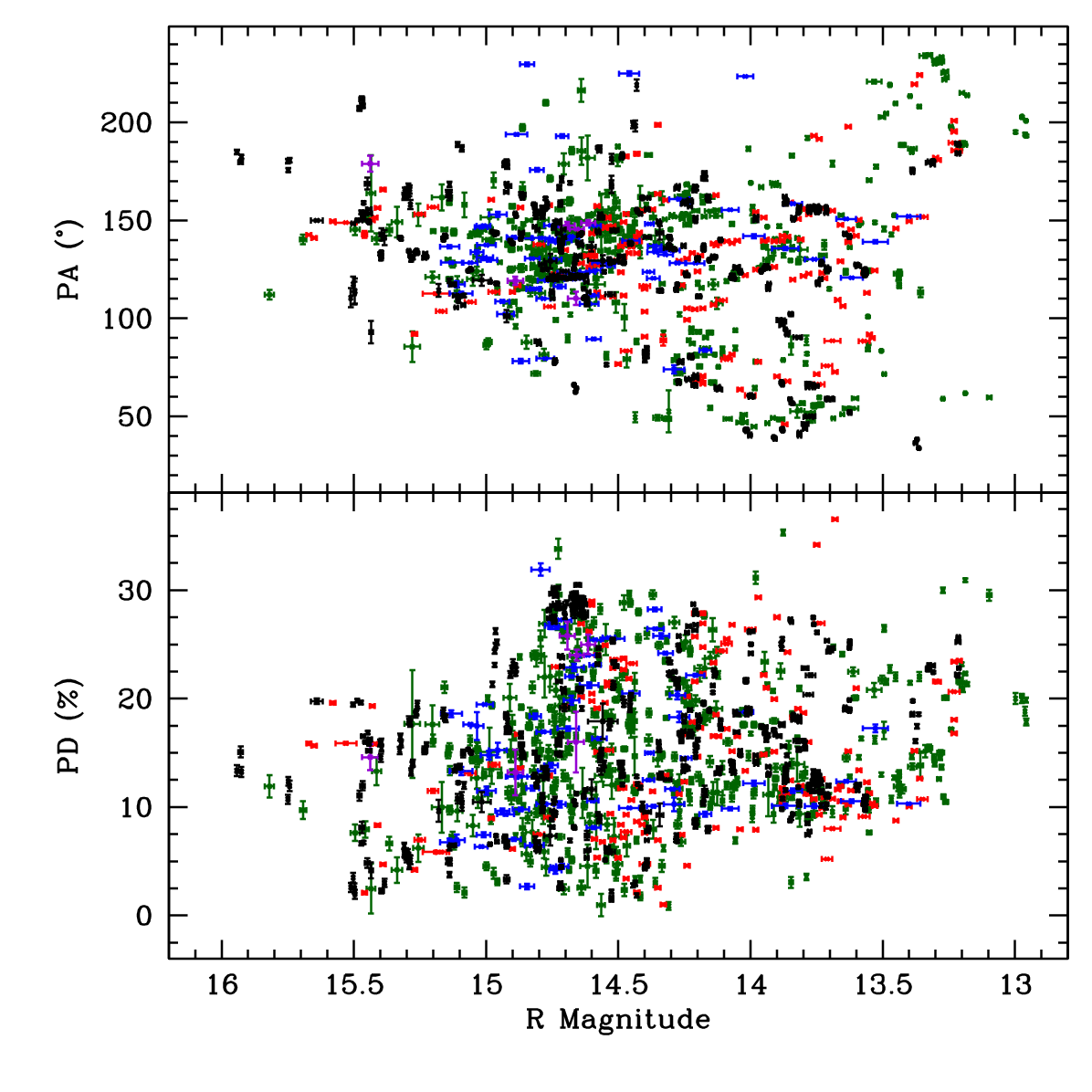}
    \caption{Variation of PD and PA with flux using the simultaneous optical flux and polarization data (within $<15$ minutes) of OJ 287 from 2015 to 2023. Symbols of different colors are the same as in Figure \ref{fig:LC}.
}
    \label{fig:magVsPDPA}
\end{figure}

\subsection{Magnitude and Polarization Variation}
\noindent
R-band optical magnitude, PD, and PA of the blazar OJ 287 are plotted in the bottom, middle, and top panels of Figure \ref{fig:LC}, respectively. Expanded versions of the same data 
are plotted for different segments of the entire data set in Figure \ref{fig:LC2}. The data plotted in Fig. \ref{fig:LC} were taken over eight consecutive observing seasons. Each observing season data is displayed as a separate segment in Figure \ref{fig:LC2}. As we noticed some peculiar variability features in Segment 3, we divided it into two sub-segments: Segment 3A and Segment 3B. \\
\\
On visual inspection, we also noticed that the magnitude and PD are correlated, in the sense that the PD rises when OJ 287 is brighter, in Segments 1 -- 3A, and Segments 4 -- 8. The general correlated PD--flux behavior is clearly apparent from Figure \ref{fig:magVsPDPA}. PA, on the other hand, appears to fluctuate around $\sim$ 140$^\circ$ - 150$^\circ$, indicating a preferential direction. A departure from this behavior, peculiar and rarely observed, is in Segment 3B, where the magnitude and PD are anti-correlated. \\  
\\
We performed statistical tests to confirm and assess the significance (p-value) of the systematic trends between R-magnitude and PD  noted by visual inspection. We  employed both the generally used Pearson-r (for apparent linear trends) and the non-parametric Kendall-tau rank correlation\footnote{\url{https://www.statisticshowto.com/kendalls-tau/}} test as it does not require normality of the input data and fares well for comparatively small data samples for the exploration of trends or relations. The p-value quantifies the null hypothesis that the data are independent, i.e., no relation between the input measurements. A p-value of $<0.05$ or $0.01$ respectively reject the null hypothesis at 5\% or 1\% levels, indicating that there is a significant relation between the two data sets. 
We give results for Pearson-r and Kendall-tau tests between PD and PA with respect to the R-band magnitude for each segment in Table \ref{tab:mag.PDPA}. For Pearson-r, the straight line fits  for PD vs R-band magnitude, and PA vs R-band magnitude values are shown in Figure \ref{fig:magVsPDPA2}. The slope of the straight line fit, $m$, in Table \ref{tab:mag.PDPA} shows the strength of the correlation or anti-correlation in PD and PA with respect to R-band brightness. \\
\\
The test results corroborate the visual trends noted above, as the p-values from both tests establish that PD is correlated ($p < 0.05$) with magnitude except for segment 3A. The test results also establish the peculiarity of segment 3B where the correlation is the opposite of the others. We note that even though the Kendall-tau test does not indicate PD vs magnitude in segment 3A are correlated, the continuity of data, along with the correlation trend of segment 2 points to 3A being the transition phase between positive and negative correlations and thus, a  crucial input. The same cannot be said for Segment 4 as there is a seasonal gap between 3B and 4. The PA vs R-magnitude results given in Table \ref{tab:mag.PDPA} are not as systematically correlated, and for many, the p-values do not reject the null hypothesis of no correlation.

\subsubsection{Segment 1 (JD 2457284 - JD 2457536)}
\noindent
Segment 1 is taken from JD 2457284 to 2457536 and is displayed in the top left panel of Figure \ref{fig:LC2}. This segment shows two flares peaking at $\sim$ JD 2457365 (2015 Dec 8) and at $\sim$ JD 2457400 (2016 Jan 12); the latter displays correlation with PD. The first peak was studied in detail by \cite{2016ApJ...819L..37V}. Here we see the declining part of the low-polarization
flare, followed by the highly polarised secondary flare. The second flare, peaking at $\sim$ JD 2457400, shows a smooth rotation of $\sim \rm{150}^{\circ}$ of PA (from $\sim \rm{200}^\circ$ to 50$^{\circ}$) \citep{2017Galax...5...83V}. Using the entire data of this segment and the Kendall-tau test, we found PD is strongly correlated with R-band brightness, whereas PA is strongly anti-correlated with it. The straight line fitting results are plotted in the top left sub-figure of Figure \ref{fig:magVsPDPA2}, and fitting parameters are provided in Table \ref{tab:mag.PDPA}.    

\subsubsection{Segment 2 (JD 2457645 - JD 2457920)}
\noindent
Segment 2, taken from JD 2457645 to 2457920, is displayed in the second panel of the top row of Figure \ref{fig:LC2}. The flux and PD exhibit  similar trends and are strongly correlated. From the beginning of the segment JD 2457645 (2016 Sep 13) to JD 2457865 (2017 Apr 21), the PA showed smooth rotation of $\sim \rm{135}^{\circ}$ (from $\sim \rm{235}^\circ$ to 100$^{\circ}$) as had already been noted by \citet{2017Galax...5...83V}. Using all these observations  we found PD and PA both are strongly correlated with R-band brightness (see the top middle sub-figure of Figure \ref{fig:magVsPDPA2}).

\subsubsection{Segment 3A (JD 2458016 - JD 2458142)}
\noindent
Segment 3A is taken to span JD 2458016 to 2458142 and is displayed in the third panel on the top row of Figure \ref{fig:LC2}. From JD 2458016 to 2458060, a gradual decrease in flux is seen whereas PD and PA show more random variations. This was the time of a major fade at 2017.85 in the OJ~287 light curve \citep{2019AJ....157...95G,2021ApJ...921...18K,2022MNRAS.514.3017V}. A peculiar behavior of flux, PD and PA are observed between JD 2458070 (2017 Nov 12) and 2458095 (2017 Dec 8) when there is a clear anti-correlation between flux and PD, whereas the flux and PD are correlated from 2458095 (2017 Dec 8) to 2458145 (2018 Jan 26), but the PA is almost non-variable in this duration. However, an examination of the entirety of these data does not show any strong correlation of PD or PA with R-band brightness because of large systematic variations within the segment (top right sub-figure of Figure \ref{fig:magVsPDPA2}). 

\subsubsection{Segment 3B (JD 2458156 - JD 2458292)}
\noindent
Segment 3B describes the after-fade behavior and it is taken from JD 2458156 to 2458292 and is shown in the top right panel of Figure \ref{fig:LC2}. A strong peculiar behavior between optical flux, PD and PA are observed in this segment from JD 2458156 (2018 Feb 6) to JD 2458270 (2018 May 31). An anti-correlation between flux and PD was present along with smooth rotations of PA of $\sim$ 30$^{\circ}$ (from 125$^{\circ}$ to 155$^{\circ}$) and then by $\sim$ 40$^{\circ}$ (from 155$^{\circ}$ to 115$^{\circ}$), from JD 2458156 to JD 2458210 and JD 2458210 to JD 2458270, respectively. Analysis of all the observations during this segment show the PD is strongly anti-correlated with R-band brightness, whereas PA does not show any correlation with R-band brightness. The straight line fitting results are plotted in the left sub-figure of the middle row of Figure \ref{fig:magVsPDPA2}, and fitting parameters are provided in Table \ref{tab:mag.PDPA}.

\subsubsection{Segment 4 (JD 2458380 - JD 2458632)}
\noindent
Segment 4 spans JD 2458380 (2018 Sep 18) to 2458632 (2019 May 28) and is displayed in the bottom left panel of Figure \ref{fig:LC2}. In this whole segment, the flux, PD and PA appear to be well correlated. 
This is a rather quiet period in OJ~287, and it precedes the 2019 July/August flare \citep{2020ApJ...894L...1L}. 
We found PD is correlated with R-band brightness, whereas PA does not seem to be correlated with R-band brightness throughout this segment (see the central panel of Figure \ref{fig:magVsPDPA2}), with fitting parameters given in Table \ref{tab:mag.PDPA}.  

\subsubsection{Segment 5 (JD 2458755 - JD 2459009)}
\noindent
Segment 5 is taken to run from JD 2458755 to 2459009 and is illustrated in the second panel of the bottom row of Figure \ref{fig:LC2}. From JD 2458755 (2019 Sep 28) to JD 2458850 (2020 Jan 1), the flux, PD and PA show random large variations.  From JD 2458850 (2020 Jan 1) to JD 2459009 (2020 Jun 8) the flux and PD are correlated while there is a smooth variation of PA of $\sim$ 130$^{\circ}$ (from 170$^{\circ}$ to 40$^{\circ}$). This is thought to be the tidal flare following the 2019 disk impact \citep{2011ApJ...742...22V}. Throughout this segment, we found PD is strongly correlated with R-band brightness, whereas PA is strongly anti-correlated with R-band brightness (right sub-figure of middle row of Figure \ref{fig:magVsPDPA2}).
    
\subsubsection{Segment 6 (JD 2459108 - JD 2459338)}
\noindent
Segment 6 is taken from JD 2459108 to 2459338 and is shown in the third panel of the bottom row of Figure \ref{fig:LC2}. From JD 2459100 (2020 Sep 7) to JD 2459178 (2020 Nov 24, in fractional years 2020.90), the flux and PD are correlated, and during this period the PA shows a smooth rotation of $\sim \rm{30}^{\circ}$ (from 150$^{\circ}$ to 120$^{\circ}$). The smooth rotation continues until JD 2459191 (2020 Dec 7, 2020.93) to about PA 80$^{\circ}$ while the flux rises, but the PD drops fast from about 20 percent to close to zero. Then the PA jumps to 162 degrees and is steady in the next stage where the both the flux and PD rise fast, the latter to around 25 percent with the rate of 6 percent per day. The peak of the flux and the PD occur at JD 2459195 (2020.95). The size of the rapid flare peaking at 2020.95 is about 4 mJy in the R-band. This coincides with the time of the expected precursor flare of the 2022 disk impact at $2020.96 \pm0.1$, and it agrees with expectation both in rapidity and size of the flare \citep{2013ApJ...764....5P}. From JD 2459300 (2021 Mar 26) to JD 2459340 (2021 May 5), the blazar rapidly faded by $\sim$ 0.8 mag (14.2 to 15.0 mag), while the PA showed a smooth rotation of $\sim$ 110$^{\circ}$ (from 120$^{\circ}$ to 230$^{\circ}$) but the PD is almost constant in the range of 3\% to 8\%. The entire data of this segment show the PD to be correlated with R-band brightness whereas there is no correlation of PA with R-band brightness, as seen in the bottom left sub-figure of Figure \ref{fig:magVsPDPA2}.

\subsubsection{Segment 7 (JD 2459520 - JD 2459750)}
\noindent
Segment 7 spans JD 2459520 to 2459750 and is displayed in the bottom second from the right panel of Figure \ref{fig:LC2}. From JD 2459575 (2021 Dec 26) to JD 2459735 (2022 Jun 4), the flux, PD and PA follow the same variability trend and are correlated. This is the period when it has been argued \citep{2023Galax..11...82V} that the disk impact was detected from three different signals: a Roche lobe flare at JD 2459531, a Gamma-ray flare at JD 2459600 and a blue flash at JD 2459635. \\
\\
In the astrophysical simulations of \citet{2013ApJ...764....5P} a strong intranight flare is produced when
the Roche lobe of the secondary is suddenly filled by the primary disk gas, leading to a strong intraday flare from the secondary \citep{2013ApJ...764....5P,2023Galax..11...82V}. Unfortunately our dense coverage of polarization (at Perkins) between JD 2459523 and JD 2459530 was temporarily halted just before this flare. This intraday flare is an order of magnitude brighter than intraday flares observed previously, and by far the largest flare of its kind ever seen in OJ~287. Because of its short duration, it is easy to miss, and was detected only by the dense total flux variability campaign by  \cite{2023MNRAS.521.6143V}. \\
\\
The Gamma-ray flare was the largest flare of its kind in OJ~287 for six years. Within the extended binary SMBH scenario including a secondary BH jet, the gamma-ray flare should arise
from the wind of the disk gas pressing on the secondary jet during the crossing of the secondary black hole through the primary disk \citep{2010A&A...522A..97A}. How this gamma-ray flare deviates from a typical blazar flare is discussed in \citet{2023Galax..11...82V}. In the optical this flare shows up moderately in total flux, but has peculiar polarization behavior: the degree of polarization falls toward the flux peak while the direction of polarization swings widely \citep{2023MNRAS.521.6143V}. Our observations add two more points (from Bulgaria) to the previously published polarization light curve. They show that taken together with the earlier data, the direction of polarization swings by 70 degrees during this flare in just 4 days. \cite{2023Galax..11...82V} associate this flare with extra emission of the secondary jet, which has different polarization properties from the primary jet and has a faster variability time scale than the primary jet \citep{2021MNRAS.503.4400D}. \\
\\
The blue flash is so named because of its exceptionally blue color. Before 2022, such a flare had been observed only once, in 2005, during a very similar crossing of the secondary black hole through accretion disk of the primary \citep{2023Galax..11...82V}. Here we are able to add 7 more points (from Crimea) to the blue flash light curve published in \cite{2023MNRAS.521.6143V} between JD 2459639 and JD 2459650. They confirm the maximum degree of polarization of $\sim 30\%$ at JD 2459643, about a week later than the flux maximum as was calculated by \citet{1971swng.conf..245V} for a flare in 3C 273, with the scaling modified for the 2022 flare in OJ 287 by \citet{2023Galax..11...82V}. \\ 
\\
The indirect after-effects of the impact such as the low-polarization flare were expected during July 2022 when OJ~287 was not observable by our instruments \citep{2023MNRAS.521.6143V,2023Galax..11...82V}. The entire data of the segment show PD and PA are correlated with R-band brightness. The straight line fitting results are plotted in the bottom middle sub-figure of Figure \ref{fig:magVsPDPA2}, and fitting parameters are provided in Table \ref{tab:mag.PDPA}.

\subsubsection{Segment 8 (JD 2459810 - JD 2459999)}
\begin{figure*}
    \centering
    \includegraphics[scale=0.9]{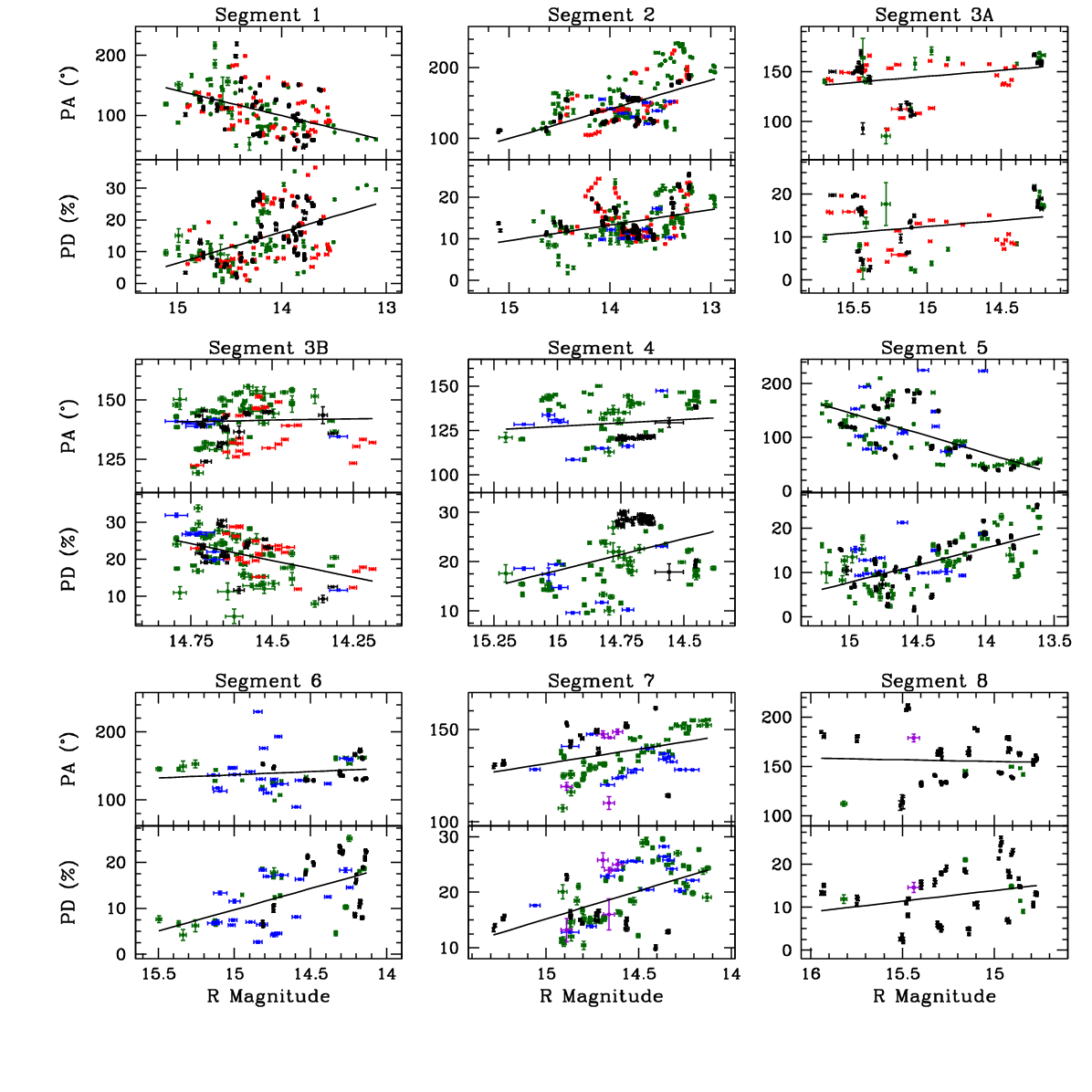}
    \caption{Variation of PD and PA with respect to R-band magnitude for different segments of light curves of OJ 287 from 2015 to 2023. Symbols of different colors are as in Figure \ref{fig:LC}.}
    \label{fig:magVsPDPA2}
\end{figure*}

\noindent
Segment 8 spans JD 2459810 to 245999 and is shown in the bottom right panel of  Figure \ref{fig:LC2}. Flux and PD follow similar trends, whereas a smooth rotation of PA of $\sim$100$^{\circ}$ (from 100$^{\circ}$ to 200$^{\circ}$) is observed. In this segment, OJ 287 was in a low flux state. The entire data of the segment show PD shows a weak correlation with R-band brightness whereas there is no correlation of PA with R-band brightness, as shown in the last sub-figure of Figure \ref{fig:magVsPDPA2} and  Table \ref{tab:mag.PDPA}.

\begin{deluxetable*}{lcrrrrrr}
\tablenum{1}
\tablecaption{PD and PA variation of OJ 287 with respect to R-band magnitude (R) throughout this campaign}
\tablewidth{0pt}
\tablehead{
\colhead{Seg.} & \colhead{Data} & \colhead{m} & \colhead{c} &
\colhead{r} & \colhead{p(r)}  & \colhead{$\tau$} & \colhead{p($\tau$)}
}
\decimalcolnumbers
\startdata
1  &  PD vs R & -9.77  & 152.97  & -0.47 & 3.5e-14 &  -0.31 &  9.7e-13 \\
   &  PA vs R & 41.64  & -483.41 & 0.42  & 2.0e-11 &   0.29 &  1.6e-11 \\
2  &  PD vs R & -3.72  & 65.33   & -0.35 & 3.0e-10 &  -0.12 &  2.5e-03 \\
   &  PA vs R & -41.28 & 719.09  & -0.51 & 3.6e-21 &  -0.32 &  6.3e-17 \\
3A &  PD vs R & -2.86  & 55.37   & -0.25 & 3.3e-02 &  -0.08 &  2.9e-01 \\
   &  PA vs R & -12.24 & 328.75  & -0.29 & 1.1e-02 &  -0.25 &  1.5e-03 \\
3B &  PD vs R & 18.09  & -242.63 & 0.44  & 4.8e-07 &   0.29 &  1.8e-06 \\
   &  PA vs R & -1.98  & 170.18  & -0.03 & 7.6e-01 &  -0.14 &  2.6e-02 \\
4  &  PD vs R & -12.79 & 210.07  & -0.37 & 2.3e-04 &  -0.25 &  3.7e-04 \\
   &  PA vs R & -7.38  & 238.21  & -0.13 & 2.2e-01 &  -0.14 &  3.9e-02 \\
5  &  PD vs R & -7.80  & 124.79  & -0.59 & 2.4e-17 &  -0.38 &  6.1e-13 \\
   &  PA vs R & 75.20  & -982.26 & 0.65  & 2.7e-21 &   0.46 &  5.6e-19 \\
6  &  PD vs R & -9.22  & 148.07  & -0.51 & 2.9e-07 &  -0.36 &  5.4e-07 \\
   &  PA vs R & -9.55  & 280.01 & -0.16 & 1.3e-01  &  -0.17 &  1.8e-02 \\
7  &  PD vs R & -10.10 & 166.64  & -0.52 & 2.1e-08 &  -0.35 &  1.7e-07 \\
   &  PA vs R & -15.75 & 367.68  & -0.36 & 2.0e-04 &  -0.26 &  1.1e-04 \\
8  &  PD vs R & -5.02  & 89.16   & -0.24 & 2.8e-02 &  -0.16 &  3.2e-02 \\
   &  PA vs R & 3.32   & 105.24  & 0.05  & 6.7e-01 &  -0.04 &  5.6e-01 \\
\enddata
\label{tab:mag.PDPA}
\tablecomments{Seg.\ = Segment; m = slope of PD or PA against R, c = intercept of PD or PA against R; r = Pearson coefﬁcient, {p(r) = its null hypothesis probability; $\tau$ = Kendall-tau coefficient,
p($\tau$) = its null hypothesis} probability.}
\end{deluxetable*}

\subsection{Position Angle Variation}
\noindent
Along with substantial changes in flux, strong and frequent PA and PD variations are generic traits of blazars, and generally the  flux and PD variations are correlated. Given that OJ 287 is an excellent binary black hole candidate it makes sense to investigate its polarization within that framework.
The electric vector polarization angle was already studied in that scenario by \citet{2012MNRAS.421.1861V}. In the model it was assumed that variations in the innermost accretion disk of the primary SMBH are translated into variations in the jet direction. In this picture the jet is assumed to line up with the rotation axis of the inner disk (disk model). The model was improved by \citet{2021MNRAS.503.4400D}, who also studied the possibility that the jet is lined up with the spin axis of the primary black hole (spin model). The disk model provides more structure in the short term, but overall the two calculations follow each other well. \\
\\
While \citet{2021MNRAS.503.4400D} compared the model only with the radio jet data, \citet{2021Galax..10....1V} extended the model to optical PA data. The jet parameters were assumed to be identical with the radio jet parameters found from fitting to observations at three different radio frequencies. This leads to a unique model for the optical PA evolution, assuming that the optical polarisation PA can be taken as proxy of jet PA angle. The model explains the overall evolution of the optical PA from 1970 up to 2018. One has to note, however, that large scatter arises in the PA during periods of major outburst activity. Thus the model should be applied only to the low states of OJ~287. This requirement is experimentally based \citep{2021Galax..10....1V}, but it could be justified theoretically by noticing that shock waves in jets may not only compress the magnetic fields in jets, but may also change the position angle of the dominant field component. Thus a period when shock-wave activity is minimal is  preferable for this model \citep{2012MNRAS.421.1861V}. \\
\\
In the present data set, the early part, from 2015 to 2017.5, is a period of major outburst activity. Therefore we have identified only short sections of the data where the activity was temporarily at a low level, and the average PA and its one-sigma scatter (SD) were calculated. Also we calculate the standard error of the mean (SEM) for each sample. These values are reported in Table \ref{tab:cal.PA} together with the expected model value from \citet{2021Galax..10....1V} for the disk model. The value for the spin model PA is about $-$40 degrees (corresponding to $+$140 degrees in our figures), and it remains constant over the period of this study. The uncertainties in the calculated PA values are about $\pm$4 degrees.
The latter part of the present data set is better suited for comparison with models. During this span the only major activity period is in the spring of 2020. We have excluded this period from our comparison. These data are similarly presented in Table \ref{tab:cal.PA}. We note that overall our data agrees better with the disk model.

\begin{deluxetable}{ccccccc}
\tablenum{2}
\tablecaption{Polarisation PA at low states of OJ287}
\tablewidth{0pt}
\tablehead{
\colhead{Start Date} & \colhead{End Date} & \colhead{No.} & \colhead{PA} &
\colhead{SD} & \colhead{SEM} & \colhead{Model} 
}
\decimalcolnumbers
\startdata
11-10-2015 & 05-11-2015 & 13 & -35 & 15 & 4 & -20\\
06-01-2016 & 13-01-2016 & 5 & -30 & 9 & 4 & -20\\
23-01-2016 & 11-02-2016 & 7 & -30 & 3 & 1 & -20\\
11-04-2017 & 19-04-2017 & 11 & -32 & 14 & 4 & -25\\
20-09-2017 & 23-06-2018 & 153 & -35 & 11 & 1 & -30\\
08-10-2018 & 14-12-2018 & 17 & -39 & 4 & 1 & -40\\
29-01-2019 & 27-05-2019 & 29 & -45 & 12 & 2& -46\\
30-09-2019 & 18-03-2020 & 33 & -50 & 44 & 8 & -50\\
17-09-2020 & 30-11-2020 & 14 & -52 & 13 & 4 & -53\\
25-09-2022 & 17-02-2023 & 84 & -25 & 25 & 3 & -50 \\
\enddata
\label{tab:cal.PA}
\tablecomments{Columns 1 and 2: beginning and the end of the data set, column 3: number of points in the set, column 4: mean position angle of the polarization vector, column 5: standard deviation of the scatter, column 6: standard error of the mean, column 7: expected model values. 180 degrees have been added to the PA values in 2016.}
\end{deluxetable}

\section{Discussion}\label{sec:discuss}
\noindent
In the present study of OJ 287, 
PD and flux are positively correlated (refer to Figure \ref{fig:magVsPDPA}) in general with almost every flux change being accompanied by a change in PD and often PA too. This is consistent with studies reporting flux and optical polarization variations in BL Lac objects which show them to be very dynamic sources in the polarization domain \citep[e.g.][]{2023MNRAS.523.4504O}. The optical flux/brightness variation of OJ 287 in terms of amplitude is similar to that seen at NIR bands over a decade duration \citep[$\sim$ 2.8 mag;][]{2022ApJS..260...39G} if the long-term base evolution is taken into account, as pointed out in \citet{2022ApJS..260...39G}. The fluctuation of PA values around $\sim$ 140$^\circ$ in the present study indicate a preferential direction and is also consistent with the result of \citet{2023MNRAS.523.4504O} using almost a decade of optical polarization study of blazars from the Steward data, except for the ambiguity of 180$^\circ$ in PA. The authors have argued such behavior can be attributed to the presence of large-scale magnetic field and turbulence, as has been argued from RoboPol studies of blazars \citep{2016MNRAS.463.3365A}.\\
\\
Flux observations at different wavelengths, as well as polarization measurements at radio and optical bands, provide valuable information for understanding the behavior of blazars and for modeling the physics of their jets. It is generally believed that the emission mechanism in blazars in the radio to optical bands is predominantly synchrotron radiation emitted by relativistic jets, enhanced by Doppler boosting. Polarimetric observations are needed to study the structure of the magnetic fields in the jets that is required to produce synchrotron radiation. \\
\\
In the present work, we observed varied combinations of flux, PD, and PA fluctuation in the various data segments presented in Figure \ref{fig:LC2}. 
Below we mention some interesting aspects of the polarization behavior with their implications for the binary black hole model of OJ~287. \\
\\
The anti-correlation between the optical flux and PD is most clear in Segment 3B in early 2018, following a strong fade in the optical flux. In the SMBH binary model this is the time when the secondary impacts the jet from its side. During this period the secondary moves from (-5000 AU, -8400 AU) to (-3100 AU, -8000 AU), in the coordinates specified in Figure 1 of \citet{2023MNRAS.tmp.2123V}. In the same figure one can see that the outline of the jet may go through the same coordinate interval. In principle this could lead to magnetic field compression which is orthogonal to the usual compression when shock waves propagate along the jet. This could be one of the explanations of why at this time the PD behaves differently from usual positive correlation with the optical flux. It is also possible that the fade in the optical flux is related to jet bending, which produces a temporarily reduced Doppler boosting just before the impact on the jet \citep{2022MNRAS.514.3017V}. \\
\\
Another interesting moment in the light curve is at the end of 2020, in Segment 6. The optical flux rises monotonically between JD 2459175 (2020 Nov 21) and JD 2459195 (2020 Dec 11), while the PD initially drops from about 20$\%$ to 5$\%$ and then rises again rapidly to about 25$\%$. Thus we see both the correlation and anti-correlation in the same flare. This is classed as a precursor of the 2022 disk impact in the binary black hole model. It is thought to arise from a superposition of emission from both the primary and the secondary jets \citep{2023Galax..11...82V}, which should definitely show up as exceptional polarization behavior.

\section{Summary}\label{sec:sum}
\noindent
We have carried out an intensive and extensive quasi-simultaneous monitoring of the optical flux and polarization of the binary super massive black hole blazar OJ 287. Our data were collected between 2015 and 2023 using optical telescopes in the USA, Japan, Crimea, Russia, and Bulgaria. We have also included public archival data from Steward Observatory, USA. The entire duration of these  observations covers eight observing seasons, and each observing season's data is treated as a separate segment. A summary of our results are as follows: 
\begin{enumerate}
\item[{\bf 1.}] Over the span of these observations, OJ 287 shows a big optical R-band flux variation of $\sim$3.0 mag, a large change in PD of $\sim$37\%, and a substantial range in PA of $\sim$215$^{\circ}$.
\item[{\bf 2.}] In general, magnitude and PD are correlated in the sense that PD rises with OJ 287 gets brighter, and vice-versa. But in the Segment 3B, the magnitude and PD are anti-correlated.
\item[{\bf 3.}] During the course of these observations, different temporal segments show various features in PA, e.g., smooth rotation, non variable, and random variation.
\item[{\bf 4.}] The detection of anti-correlation in optical flux and PD in Segment 3B is a rare finding in OJ 287. In this segment a smooth rotation of PA of $\sim$ 30$^{\circ}$ (from 125$^{\circ}$ to 155$^{\circ}$) is followed by an opposite rotation of $\sim$ 40$^{\circ}$ (from 155$^{\circ}$ to 115$^{\circ}$), from JD 2458156 to JD 2458210 and JD 2458210 to JD 2458270, respectively.
\item[{\bf 5.}] Another exceptional finding is at the end of 2020 when both correlated and anti-correlated behavior occurs in the same flare. This is the time when in the binary model the secondary jet was expected to show up and overpower the primary jet for a short period of time \citep{2023Galax..11...82V}. This may be the first time (through polarization at least) that we actually see the emission of the secondary black hole in the binary black hole system of OJ~287. Overall, the PA of polarization follows the binary model during the low brightness states of OJ~287 throughout the entire period of these observations. 
\end{enumerate}

\section*{ACKNOWLEDGMENTS}
\noindent
Data from the Steward Observatory spectropolarimetric monitoring project were used. This program is supported by Fermi Guest Investigator grants NNX08AW56G, NNX09AU10G, NNX12AO93G and NNX15AU81G. This study was based in part on observations conducted using the Perkins Telescope Observatory (PTO) in Arizona, USA, which is owned and operated by Boston University. \\
\\
We thank the anonymous reviewer for useful comments. ACG is partially supported by Chinese Academy of Sciences (CAS) President's International Fellowship Initiative (PIFI) (grant no. 2016VMB073). PK acknowledges support from the Department of Science and Technology (DST), government of India, through the DST-INSPIRE Faculty grant (DST/INSPIRE/04/2020/002586). RI's work is supported by JST, the establishment of university fellowships towards the creation of science technology innovation, Grant Number JPMJFS2129. MFG is supported by the National Science Foundation of China (grant 11873073), Shanghai Pilot Program for Basic Research-Chinese Academy of Science, Shanghai Branch (JCYJ-SHFY-2021-013), the National SKA Program of China (Grant No. 2022SKA0120102), the science research grants from the China Manned Space Project with No. CMSCSST-2021-A06, and the Original Innovation Program of the Chinese Academy of Sciences (E085021002). The research at Boston University was supported in part by National Science Foundation grant AST-2108622, and a number of NASA Fermi Guest Investigator grants; the latest is 80NSSC22K1571. ZZ is thankful for support from the National Natural Science Foundation of China (grant no. 12233005). RB and AS are partially supported by the Bulgarian National Science Fund of the Ministry of Education and Science under grants KP-06-H38/4 (2019), KP-06-KITAJ/2 (2020), and KP-06-H68/4 (2022). HG acknowledges the financial support from the DST, GoI, through DST-INSPIRE faculty award IFA17-PH197 at ARIES, Nainital, India.

\bibliography{ref}{}

\bibliographystyle{aasjournal}

\end{document}